\documentclass[aps,prb,twocolumn,showpacs,superscriptaddress,amsmath,amssymb]{revtex4-1}

\usepackage[pdftex]{graphicx}
\usepackage[pdftex,bookmarks=true,bookmarksopen,bookmarksnumbered,
                colorlinks,
                linkcolor=blue,
                citecolor=blue]{hyperref}
\usepackage{dcolumn}
\usepackage{bm}

\begin{document}


\title{A Waveguide-Coupled On-Chip Single Photon Source}

\author{A. Laucht}
\affiliation{Walter Schottky Institut and Physikdepartment, Technische Universit\"at M\"unchen, Am Coulombwall 4a, 85748 Garching, Germany}%
\affiliation{Centre for Quantum Computation \& Communication Technology, The University of New South Wales, Sydney NSW 2052, Australia}
\author{S. P\"utz}
\affiliation{Walter Schottky Institut and Physikdepartment, Technische Universit\"at M\"unchen, Am Coulombwall 4a, 85748 Garching, Germany}%
\author{T. G\"unthner}
\affiliation{Walter Schottky Institut and Physikdepartment, Technische Universit\"at M\"unchen, Am Coulombwall 4a, 85748 Garching, Germany}%
\affiliation{Institute f\"ur Experimentalphysik, Universit\"at Innsbruck, Technikerstrasse 25, 6020 Innsbruck, Austria}
\author{N. Hauke}
\affiliation{Walter Schottky Institut and Physikdepartment, Technische Universit\"at M\"unchen, Am Coulombwall 4a, 85748 Garching, Germany}%
\author{R. Saive}
\affiliation{Walter Schottky Institut and Physikdepartment, Technische Universit\"at M\"unchen, Am Coulombwall 4a, 85748 Garching, Germany}%
\author{S. Fr\'ed\'erick}
\affiliation{Walter Schottky Institut and Physikdepartment, Technische Universit\"at M\"unchen, Am Coulombwall 4a, 85748 Garching, Germany}%
\affiliation{Institute for Microstructural Sciences - National Research Council of Canada, Ottawa, ON, Canada}
\author{M. Bichler}
\affiliation{Walter Schottky Institut and Physikdepartment, Technische Universit\"at M\"unchen, Am Coulombwall 4a, 85748 Garching, Germany}%
\author{M.-C. Amann}
\affiliation{Walter Schottky Institut and Physikdepartment, Technische Universit\"at M\"unchen, Am Coulombwall 4a, 85748 Garching, Germany}%
\author{A. W. Holleitner}
\affiliation{Walter Schottky Institut and Physikdepartment, Technische Universit\"at M\"unchen, Am Coulombwall 4a, 85748 Garching, Germany}%
\author{M. Kaniber}
\affiliation{Walter Schottky Institut and Physikdepartment, Technische Universit\"at M\"unchen, Am Coulombwall 4a, 85748 Garching, Germany}%
\author{J. J. Finley}
\affiliation{Walter Schottky Institut and Physikdepartment, Technische Universit\"at M\"unchen, Am Coulombwall 4a, 85748 Garching, Germany}%
\email{finley@wsi.tum.de}

\date{September 26, 2011}

\begin{abstract}
We investigate single photon generation from individual self-assembled InGaAs quantum dots coupled to the guided optical mode of a GaAs photonic crystal waveguide. By performing confocal microscopy measurements on single dots positioned within the waveguide, we locate their positions with a precision better than $0.5$~$\mu$m.
Time-resolved photoluminescence and photon autocorrelation measurements are used to prove the single photon character of the emission into the propagating waveguide mode. The results obtained demonstrate that such nanostructures can be used to realize an on-chip, highly directed single photon source with single mode spontaneous emision coupling efficiencies in excess of $\beta_\Gamma\sim85$~$\%$ and the potential to reach maximum emission rates $>1$~GHz.
\end{abstract}

\pacs{42.50.Ct, 42.70.Qs, 78.67.Hc, 78.47.-p, 42.82.Et}
\keywords{quantum dot, photonic crystal, waveguide, single photon emission}
\maketitle

The ability to control the direction and rate of spontaneous emission by tailoring the local density of photon modes experienced by an emitter is a key concept to enhance the efficiency of nanoscale light sources such as single photon sources~\cite{Purcell46, Pelton02, Santori02, Claudon10, Fujiwara11, Davanco11} and nanoscale lasers.~\cite{Altug06} Over recent years, several groups have demonstrated the ability to control light-matter coupling using photonic crystal nanostructures.~\cite{Englund05, Badolato05, Kress05a} Strong enhancements of spontaneous emission rates have been observed for individual quantum emitters in low mode volume, high-Q defect cavities.~\cite{Englund05, Kaniber07, Englund10} However, for such systems to be useful one has to spectrally bring the emitter and cavity mode into mutual resonance calling for sophisticated electro-~\cite{Laucht09} or thermo-optical~\cite{Englund07} tuning methods. Recently, Viasnoff-Schwoob et al.~\cite{Viasnoff05} demonstrated that enhanced light-matter coupling can be obtained over wider bandwidths by coupling emitters to the enhanced density of photonic modes close zero-group-velocity points of the dispersion of a 1D photonic crystal waveguide.~\cite{Vlasov05} Therefore, such 1D photonic crystal waveguides (PWGs) provide strong promise for use as an on-chip single-photon source since they effectively funnel spontaneous emission into the guided optical mode and obviate the need for precise spectral tuning of the emitter - photonic system.~\cite{Hughes04,Viasnoff05,Lecamp07, Rao07, Rao07b, Rao08, Lund-Hansen08, Patterson09, Dewhurst10, Thyrrestrup10, Yao10b} A highly efficient single-photon source is the key component required in many quantum communication protocols~\cite{Bennett84} and the combination of single quantum dots (QDs) coupled to propagating modes on a photonic crystal chip is of strong interest for chip based implementations of linear optics quantum computing.~\cite{OBrien09}


In this paper, we present experimental investigations of the emission characteristics of single self-assembled InGaAs QDs coupled to the guided mode of a linear defect (W1) PWG.~\cite{Loncar00,Olivier01} We perform spatially resolved photoluminescence (PL) measurements to locate the position of the QD inside the PWG. By comparing the emission intensity and spontaneous emission dynamics detected along an axis perpendicular to the sample surface with similar measurements detected in the plane of the photonic crystal waveguide, we obtain strong evidence for significant enhancements of the radiative coupling to the propagating waveguide mode. Most notably, a $\sim55\pm8\times$ more efficient coupling to the PWG mode is measured compared to radiation into free space modes along the vertical detection axis. Time-resolved photoluminescence measurements detected on the same QD transition allow us to estimate the fraction of all spontaneous emission emitted into the waveguide mode ($\beta_\Gamma$). This can be very high for individual QDs in PWG structures,~\cite{Lund-Hansen08, Lecamp07, Rao07} and our measurements reveal a lifetime of $\tau=0.87\pm0.15$~ns from which we estimate that $85$~$\%<\beta_{\Gamma}<96$~$\%$.
Second order photon autocorrelation $g^{(2)}(\tau)$ measurements confirm the single photon character of the QD emission into the waveguide mode with a multiphoton probability of $g^{(2)}(0)=0.27\pm0.07$, compared to a Poissonian source with the same average intensity. The wide bandwidth of the PWG guided modes ($>25$~meV) provides a highly attractive route towards the design of on-chip quantum optics experiments obviating the need to fine-tune the QD transition into spectral resonance with a high-Q photonic crystal cavity mode.~\cite{Mosor05, Laucht09}

\begin{figure}[t!]
\includegraphics[width=1\columnwidth]{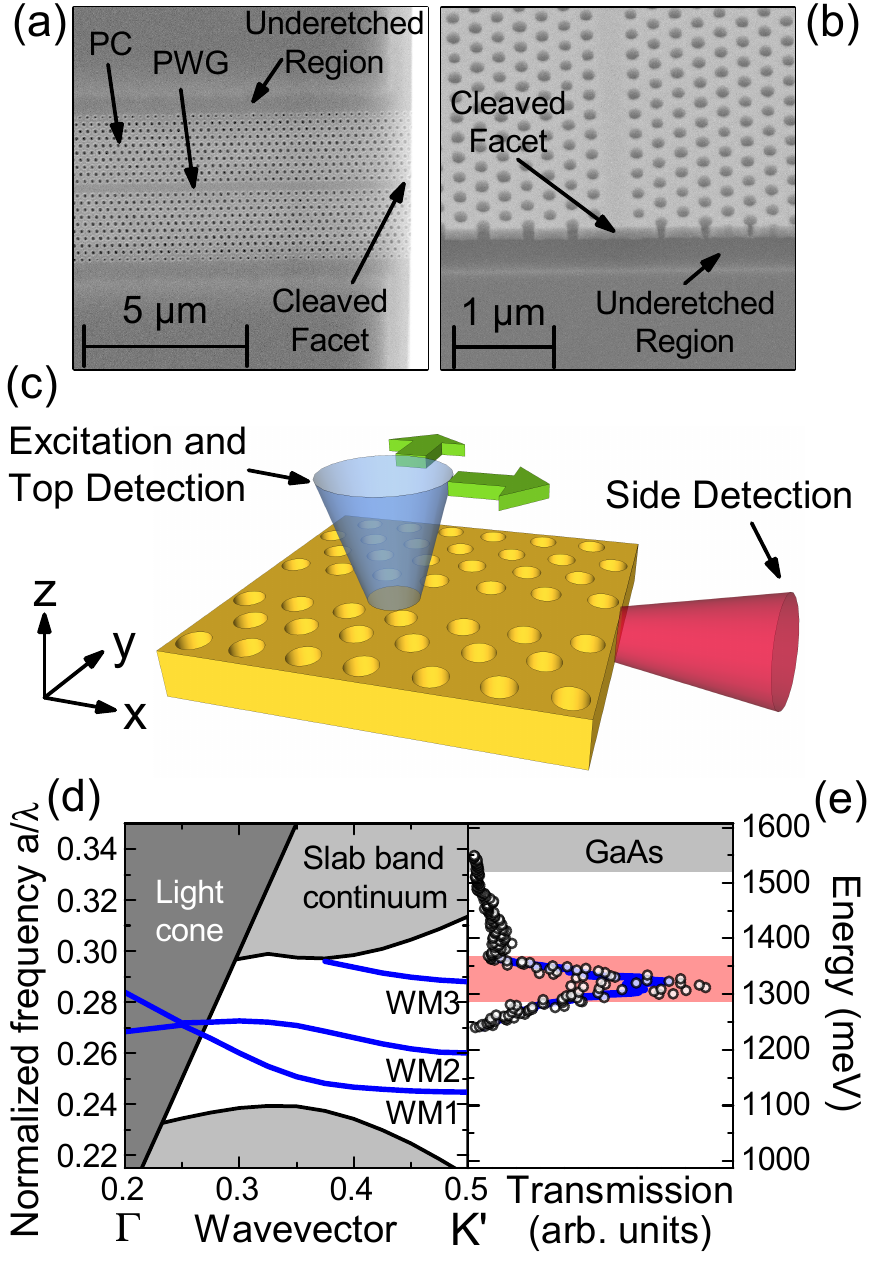}
\caption{\label{figure01} (a) and (b) SEM images of the W1 waveguide from the top and from the side, respectively. (c) Schematic of the excitation and detection geometry. (d) Photonic bandstructure calculations for a W1 waveguide with $r/a=0.26$ and $h/a=0.6667$. The solid blue lines correspond to the photonic waveguide modes and the light gray region to the slab waveguide modes. The dark gray region indicates the light cone. (e) Spectral transmission of the waveguide for illumination from the top at the inner end of the waveguide and detection from the side at the cleaved end. The red (light gray) shaded region marks the spectral region of quantum dot emission (bulk GaAs).}
\end{figure}

The sample investigated was grown by molecular beam epitaxy and consists of a $500$~nm thick Al$_{0.8}$Ga$_{0.2}$As sacrificial layer, and a $180$~nm thick GaAs layer containing a single layer of nominally In$_{0.5}$Ga$_{0.5}$As QDs at its midpoint.
The QD layer has a relatively low density $\rho_{QD}<1$~$\mu$m$^{-2}$, which allows us to selectively excite and study the emission characteristics of individual QDs using a confocal microscope in which the laser is focused to a spot with a diameter of $\sim 1.2$~$\mu$m. A two-dimensional photonic crystal formed by defining a triangular array of air holes ($r\sim71\pm3$~nm) with a nominal lattice constant of $a=270$~nm was realized using a combination of electron-beam lithography and reactive ion etching. PWGs were formed by introducing line defects consisting of a single missing row of holes (W1 waveguide).~\cite{Loncar00,Olivier01} For these specific geometrical parameters the guided waveguide modes span the energy range of $E=1125-1364$~meV ($a/\lambda=0.245-0.297$) as determined by finite-difference time-domain simulations and transmission measurements performed on reference samples [cf. Fig.\ref{figure01}~(d) and (e)].
The sample was then cleaved perpendicular to the axis of the PWG, to facilitate direct optical access to the waveguide mode and to allow for collection of light directly propagating through the waveguide. Cleaving of the sample was done before we fabricated the free-standing membrane in the wet-etching step with hydrofluoric acid. The remote end of the $45$~$\mu$m ($\sim$165 unit cells) long waveguide is terminated with an input coupler that serves to scatter light into the guided waveguide modes for transmission experiments.~\footnote{The low reflectivities of the waveguide terminations combined with the expected significant propagation losses for the third waveguide mode (WM3) are not expected to significantly modify the waveguide mode dispersion compared to the infinite waveguide approximation.~\cite{Rao07b, Rao08}}

In Fig.\ref{figure01}~(a) and (b) we present scanning electron microscope (SEM) images of the investigated sample. Fig.\ref{figure01}~(a) shows an image recorded normal to the sample surface. It shows the W1 PWG and the cleaved edge at which it ends. The cleaved facet runs perpendicular through the omitted row of air holes that define the waveguide. This can be quite easily realized when orienting the photonic structure and the cleaved edge along the [110] crystal axes of the GaAs substrate. The image also shows the underetched region of the waveguide which extends $\sim0.6$~$\mu$m from the photonic crystal into the unpatterned region of the sample. In Fig.\ref{figure01}~(b) we present an image of the cleaved facet of the sample, recorded along an axis $45^\circ$ to the waveguide axis. We can identify the smooth surface of the cleaved facet and also the free-standing membrane containing the InGaAs QDs.

For optical characterization the sample is mounted in a liquid He-flow cryostat and cooled to $T = 15$~K. For excitation we use a pulsed Ti-Sapphire laser ($80$~MHz repetition frequency, $5$~ps pulse duration) tuned to the low energy absorption edge of the bulk GaAs ($\lambda_{laser}=815$~nm). While the sample is always excited from the top (i.e. perpendicular to the sample surface) using a $100\times$ microscope objective (NA=0.50), the PL signal is either detected from the top using the same objective, or from the cleaved facet of the PWG (i.e. perpendicular to the cleaved edge) using a $50\times$ microscope objective (NA=0.42). A schematic representation of the excitation and detection scheme is shown in Fig.\ref{figure01}(c). The QD PL is spectrally analyzed using a $0.5$~m imaging monochromator and detected using a Si-based, liquid nitrogen cooled CCD detector. For time-resolved spectroscopy we use a Si-based avalanche photodiode connected to the side-exit of our monochromator with a temporal resolution of $\sim350$~ps ($\sim150$~ps after deconvolution), and for autocorrelation experiments a pair of identical detectors with a temporal resolution of $\sim750$~ps.
\begin{figure*}[t!]
\includegraphics[width=1\textwidth]{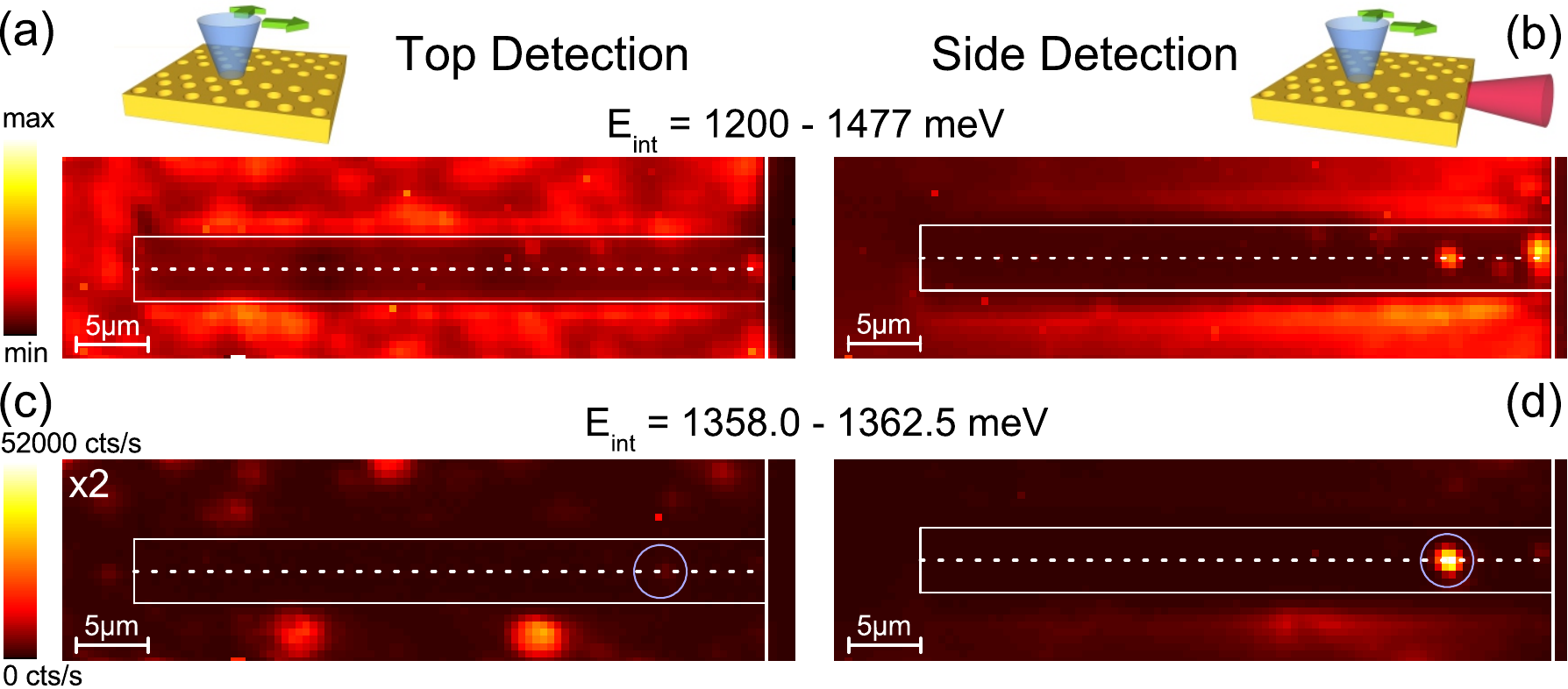}
\caption{\label{figure02} Comparison of the PL intensity for top and side detection. Spatially-resolved PL scan performed with (a),(c) detection from the top, and (b),(d) detection from the cleaved facet of the waveguide. The PL intensity is integrated over the spectral range of the wetting layer and quantum dot emission $E_{int}=1200-1477$~meV in (a) and (b), and over the limited spectral range of a single quantum dot $E_{int}=1358.0-1362.5$~meV in (c) and (d). The dotted white lines mark the position of the PWG, while the plain white lines indicate the outline the photonic crystal structure and the cleaved edge of the sample.}
\end{figure*}

We determine the spectral properties of the guided waveguide mode by conducting photonic bandstructure calculations using the software package RSoft.~\cite{rsoft} We use optical constants that are appropriate for GaAs ($n_{GaAs}=3.5$) and the geometric parameters for the investigated W1 photonic crystal waveguide ($h/a=0.6667$ and $r/a=0.26$). The result of this simulation is presented in Fig.~\ref{figure01}(d) where we plot the normalized frequency of the photonic bands as a function of k-vector along the $\Gamma$ - $K'$ point direction.~\cite{Johnson00,Dorfner08} The guided photonic waveguide are depicted as blue solid lines, the slab waveguides modes as a light gray-shaded region and the region above the light cone is shaded dark gray.~\footnote{The region above the light cone corresponds to the energy-wavevector combinations for which photons are not confined to the slab by total internal reflection, i.e. they can leave the waveguide in vertical directions.} We calculate the lowest energy waveguide mode WM1 to span the normalized frequency range $a/\lambda=0.245-0.266$, corresponding to an energy of $E=1125-1221$~meV. The second waveguide mode WM2 is at $a/\lambda=0.260-0.272$ ($E=1194-1249$~meV) and the third waveguide mode WM3 at $a/\lambda=0.288-0.297$ ($E=1322-1364$~meV).~\footnote{Due to the large unit cell used for the simulations, the third waveguide mode was hidden within folded bands for $r/a=0.26$. We, therefore, linearly extrapolated its position from six calculations performed for $r/a=0.29-0.34$.}

In order to verify the calculations and to check that the quantum dot emission is spectrally in resonance with one of the guided modes of the photonic crystal waveguide, we measure the spectral transmission of the structure. To do this, the laser is focussed on the inner end of the photonic crystal waveguide within the body of the photonic crystal. In this geometry laser light is scattered into the waveguide and the transmitted intensity can be detected from the cleaved facet as the wavelength is scanned. While this type of measurement allows us to locate the transmission band of the PWG, it is not possible to reliably estimate the transmission losses since the in-coupling and out-coupling efficiencies are unknown and difficult to reliably determine. In particular, the outcoupling efficiency depends critically on the position within the unit cell where the waveguide is cleaved. Fig.~\ref{figure01}(e) shows the spectrum of the transmitted light recorded using this method. The open circles correspond to the experimental data points while the blue solid line is a moving 7-point average. The transmission band is centered at $1323$~meV, with a width of $55$~meV. This is in fairly good accord with the simulations conducted in Fig.~\ref{figure01}(d). The gray shaded region in Fig.~\ref{figure01}(e) indicates the spectral range over which we expect absorption from the bulk GaAs, while the red shaded region indicates the range for which we observe photoluminescence emission from the quantum dots, clearly showing that they are located inside the transmission band of the photonic crystal waveguide.

We performed spatially resolved photoluminescence measurements of the PWG region by scanning the excitation spot over the sample surface and recording spectra on a $40\times15$~$\mu$m$^2$ square grid with a $0.5$~$\mu$m pitch. In Fig.~\ref{figure02}(a) and (b) we present spatially-resolved contour plots of the photoluminescence signal integrated over the spectral range $E_{int}=1200-1477$~meV, i.e. including photoluminescence from the wetting layer and the quantum dots. We performed the measurement in Fig.~\ref{figure02}(a) with excitation and detection from the top and the measurement in Fig.~\ref{figure02}(b) with excitation from the top and detection via the cleaved facet of the photonic crystal waveguide. Integration over this energy range allows us to precisely locate the position of the PWG and the cleaved facet on the luminescence maps (indicated by the solid white lines). While the PL intensity on the unprocessed material is homogeneous for top detection, we observe a general trend to higher intensities in side detection (red color) at positions closer to the cleaved facet of the waveguide, probably due to the proximity of the excitation spot to the outcoupling facet.
%
\begin{figure*}[t!]
\includegraphics[width=1\textwidth]{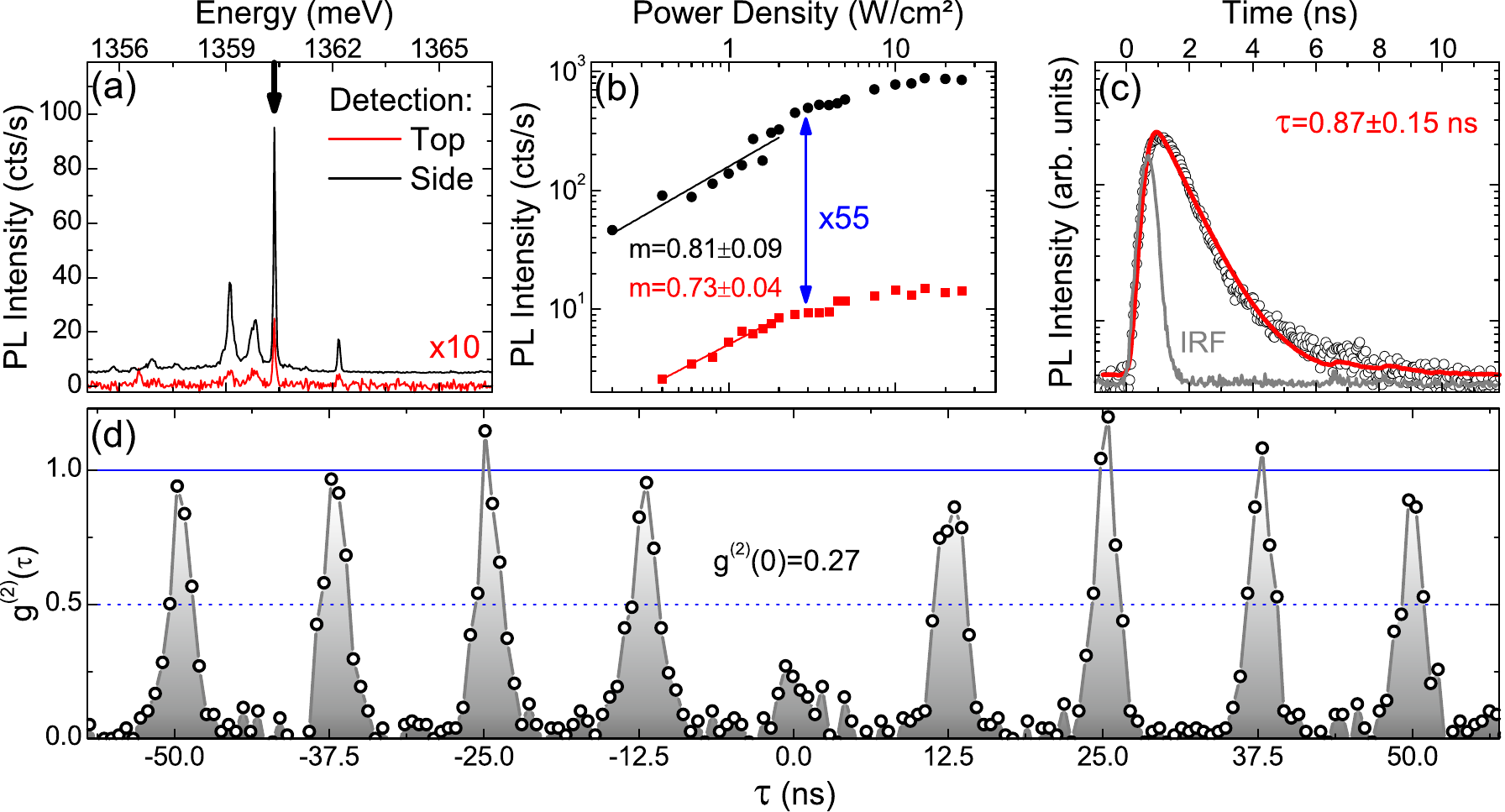}
\caption{\label{figure03} (a) PL spectra of the single quantum dot recorded with detection from the top (black line) and from the side (red line). (b) Power-dependent PL intensity of the single exciton line at $E_{X}=1360.4$~meV with detection from the top (black circles) and from the side (red squares). (c) Corresponding time-resolved PL intensity measurements with detection from the side (open circles). The solid red line is a fit to the data and solid gray line is the instrument response function of the excitation and detection system. (d) Corresponding photon autocorrelation measurement with detection from the side, proving the single photon character of the emission. The excitation power density for this measurement was $4$~W/cm$^2$.}
\end{figure*}

While integration over a wide spectral range enables us to locate the photonic crystal waveguides, integration over a narrow spectral window allows us to locate the position of individual self-assembled QDs. In Fig.~\ref{figure02}(c) and (d) we integrate the same dataset presented in Fig.~\ref{figure02}(a) and (b) over a limited spectral range $E_{int}=1358.0-1362.5$~meV. Fig.~\ref{figure02}(c) and (d) show the signal detected from the top and from the side, respectively. One particular quantum dot with an energy at the high energy side of the transmission band, at a position $d_{QD}^{edge}=7.3\pm0.5$~$\mu$m away from the cleaved facet (highlighted by the blue circle), exhibits extremely weak out-of-plane emission but comparatively much stronger in-plane emission. This observation provides evidence for good spatial coupling between this particular quantum dot and the photonic crystal waveguide mode.

In Fig.~\ref{figure03}(a) we compare photoluminescence spectra recorded for the two different detection geometries where the spectrum detected from the top is plotted as red line (note the $\times10$ enhanced scale) and detection from the side is plotted as black line. We observe the same transition lines in both detection geometries, albeit with a much lower intensity for top detection as discussed above. For the line marked with the black arrow, we conduct power-dependent measurements and plot the peak intensity in Fig.~\ref{figure03}(b) for both top (red squares) and side (black circles) detection. In both detection geometries we observe a slightly sublinear power law dependence $I\propto P^{m}$ with $m_{top}=0.73\pm0.04$ and $m_{side}=0.81\pm0.09$ for non-resonant excitation. This provides evidence that the transition line investigated has single exciton character.~\cite{Finley01, Kaniber08a, abbarchi09, Laucht10b} While the power-dependence and the onset of saturation ($P\sim8$~W/cm$^2$) is very similar for both detection geometries, at $P=3$~W/cm$^2$ the absolute detected intensity for side detection is $\sim55\pm8\times$ higher than for top detection ($I_{side}=494\pm40$~cts/s cf. $I_{top}=9\pm1$~cts/s). The intensities obtained for the two different detection geometries can be directly compared to obtain information about the relative coupling strength of the QD to the PWG mode, compared to other radiation modes of the system. Top detection of the emission provides information about the radiative emission into non-guided modes, while side detection supplies information about the radiative emission into the photonic crystal waveguide mode. A $\sim55\pm8\times$ higher intensity for side detection is, therefore, a clear signature of the efficient coupling of the QD to the propagating PWG mode.

Further support for this conclusion is provided by the time-resolved measurement presented in Fig.~\ref{figure03}(c). The black circles correspond to the measured decay transient, while the red line is a fit to the data taking into account the instrument response function (IRF) of the detection and excitation systems (gray line). For the specific transition under study we measure a lifetime of $\tau=0.87\pm0.15$~ns. From this value we obtain the $\beta_{\Gamma}$-factor which is calculated from the spontaneous emission rates into the waveguide mode ($\Gamma_{WG}$) and non-guided, radiative modes ($\Gamma_{rad}$), and the non-radiative decay rate ($\Gamma_{nr}$),~\cite{Lecamp07, Rao07, Lund-Hansen08} using the equation
\begin{equation}
\beta_{\Gamma}=\frac{\Gamma_{WG}}{\Gamma_{WG}+\Gamma_{rad}+\Gamma_{nr}}=\frac{\Gamma_{WG}}{\Gamma_{WG}+\Gamma_{int}}.
\label{betag}
\end{equation}
Here, $\Gamma_{rad}$ and $\Gamma_{nr}$ can be combined to $\Gamma_{int}=\Gamma_{rad}+\Gamma_{nr}$ which is the intrinsic emission rate of an uncoupled QD. Typical intrinsic lifetimes of reference QDs emitting into the two-dimensional photonic bandgap were measured to be $\tau_{int}=1/\Gamma_{int}\sim5-20$~ns\footnote{For comparison, the typical lifetime of the InGaAs QDs in the unprocessed GaAs is $0.5$ - $0.8$~ns.} depending on the position of the quantum dot within the photonic crystal.~\cite{Kaniber07, Kaniber08a, Lund-Hansen08} This value is much longer than the short lifetime of $\tau=0.87\pm0.15$~ns for the waveguide coupled dot. This observation indicates that the $\beta$-factor is in the range $85$~$\%<\beta_{\Gamma}<96$~$\%$, in good agreement with values from the literature.~\cite{Rao07, Lecamp07, Thyrrestrup10} Similar findings were observed for a number of different QDs within the waveguide region, indicating that the fraction of light emitted into the waveguide mode varies from $\beta<10$~$\%$ to $\beta>90$~$\%$ depending on the exact position and frequency of the dot. We note that high $\beta$-factors (i.e. efficient QD emission into the waveguide mode) can be obtained for a wide range of detunings, however high Purcell-factors~\cite{Purcell46} (i.e. emission enhancement compared to QDs in unprocessed GaAs material) are usually only obtained when the QD transition is in resonance with the flat part of the PWG mode in the dispersion relation (slow-light regime).~\cite{Lecamp07, Yao10b} Our approach allows the design of efficient broad-band single-photon sources, while the design of highly coherent single photon sources would still require a tuning mechanism to obtain large Purcell-factors.

We continue to present a photon autocorrelation ($g^{(2)}(\tau)$) measurement recorded with side detection in Fig.~\ref{figure03}(d) using pulsed excitation with a time-averaged power density of $4$~W/cm$^2$. The peak at $\tau=0$ is, clearly, much weaker than the adjacent peaks and from comparing the areas of the peaks we estimate $g^{(2)}(0)=0.27\pm0.07$. This value is significantly below the value of $0.50$, proving that almost all of the detected light originates from the investigated single exciton transition and that we observe clean single photon emission. Thus, the results presented in Fig.~\ref{figure03} demonstrate a highly directed and efficient single photon turnstile device with $\beta>85$~$\%$ and a radiative lifetime that would facilitate a maximum repetition rate $f>1$~GHz. Photons are emitted directly into an on-chip photonic waveguide providing significant flexibility for on-chip quantum optics experiments. When compared to geometries where the quantum dot is resonantly coupled to a cavity mode, in this system there is no need to spectrally tune the quantum dot into resonance and, thus, efficient single photon emission can be realized over a wide bandwidth. This opens up perspectives for quantum information experiments with wavelength division multiplexing capabilities.~\cite{treiber09}

In conclusion we investigated the spontaneous emission properties of a single self-assembled InGaAs quantum dot coupled to the mode of a photonic crystal W1 defect waveguide. We located a well-coupled quantum dot $7.3\pm0.5$~$\mu$m away from the cleaved facet of the waveguide by performing spatially resolved photoluminescence scans. When comparing the signal obtained in the two different detection geometries we observed the same spectral features for detection perpendicular to the sample surface and for detection at the cleaved end of the waveguide, albeit with a different intensity depending mainly on the coupling strength between quantum dot and waveguide mode. We estimated a $\beta$-factor of $\beta_\Gamma>85\%$ from lifetime measurements, and demonstrate efficient single photon emission with $g^{(2)}(0)=0.27\pm0.07$, making these structures prospective candidates for on-chip quantum communication and quantum optical investigations.  

We gratefully acknowledge financial support of the DFG via the SFB 631, the German Excellence Initiative via NIM, the EU-FP7 via SOLID, and the BMBF via QuaHLRep project 01BQ1036. AL acknowledges support of the TUM-GS, and SF of the Alexander von Humboldt Foundation.

\emph{Note added:} After submission of this work, another demonstration of on-chip single-photon emission of a quantum dot coupled to a photonic crystal waveguide was reported.~\cite{Schwagmann11} Our work and this work were performed independently.

\bibliography{Papers}

\end{document}